\input  phyzzx
\input epsf
\overfullrule=0pt
\hsize=6.5truein
\vsize=9.0truein
\voffset=-0.1truein
\hoffset=-0.1truein

%
%

\def\IC{{\ \hbox{{\rm I}\kern-.6em\hbox{\bf C}}}}
\def\IR{{\hbox{{\rm I}\kern-.2em\hbox{\rm R}}}}
\def\IZ{{\hbox{{\rm Z}\kern-.4em\hbox{\rm Z}}}}
\def\pa{\partial}
\def\sIR{{\hbox{{\sevenrm I}\kern-.2em\hbox{\sevenrm R}}}}

\def\Sc{Schwarzschild}
\def\bh{black hole}
%
%
\hyphenation{Min-kow-ski}

\rightline{SU-ITP-96-40}
\rightline{September  1996}
\rightline{hep-th/9609075}

\vfill

%
%
\title{\bf Counting Schwarzschild and Charged Black Holes}

\vfill

%
%
\author{Edi Halyo$^1$\foot{halyo@dormouse.stanford.edu}, Barak
Kol$^1$\foot{barak@leland.stanford.edu}, Arvind
Rajaraman$^2$\foot{arvindra@dormouse.
stanford.edu} and
Leonard Susskind$^1$\foot{susskind@dormouse.stanford.edu}}

\vfill

\address{$^1$Department of Physics,  Stanford University\break Stanford, CA
94305-4060}
\address{$^2$Stanford Linear Accelerator Center,  Stanford University\break
Stanford, CA 94305}

\vfill

%
%
We review the arguments that fundamental string states are in one to one
correspondence
with black hole states. We demonstrate the power of the assumption by showing
that it implies that the
statistical entropy of a wide class of nonextreme black holes occurring in
string theory
is proportional to the horizon area. However, the numerical coefficient
relating the area and
entropy only agrees with the Bekenstein--Hawking formula if the central charge
of the string
is six which does not correspond to any known string theory. Unlike the current
D-brane
methods the method used in this paper is applicable for the case of
Schwarzschild and highly non-extreme charged black holes.

\vfill\endpage

%
%

\REF\spec{L.~Susskind,
Rutgers University preprint RU-93-44, August 1993, hep-th/9309145.}

\REF\sen{A.~Sen  \journal Mod. Phys. Lett. & A10 (95) 2081, hep-th/9504147.}

\REF\stvaf{A.~Strominger and C.~Vafa ,
hep-th/9601029.}

\REF\candj {C.~Callan, and J.~Maldacena,
 hep-th/9602043.}

\REF\HorSt{ G.~Horowitz and A.~Strominger, hep-th/9602051.}

\REF\HorMldStr{G.~Horowitz, J.~Maldacena and A.~Strominger, hep-th/9603019.}

\REF\Mld{ J.~Maldacena, private communication.}

\REF\Unruh{W.G.~Unruh \journal Phys.Rev. & D14 (76)3251. }

\REF\Gib{G.W.~Gibbons \journal Comm. Math. Phys. & 44(75) 245.}

\REF\GibDM{S.R.~Das,G.~Gibbons, and S.D.~Mathur, hep-th/9609052.}

\REF\DM{S.R.~Das and S.D.~Mathur,hep-th/9606185.}

\REF\DMM{S.R.~Das and S.D.~Mathur,hep-th/9607149.}

\REF\DMW{A.~Dhar, G.~Mandal, and S.R.~Wadia, hep-th/9605234.}

\REF\Ford{L.H.Ford,\journal Phys.Rev. & D12(75)2963}

\REF\GSW{M.~Green,J.~Schwarz, and E.~Witten, {\it Superstring Theory},vol.1,
Cambridge
University Press(1987)}


%
%

%
%
\chapter{Introduction}

A number of years ago one of us speculated [\spec] that the statistical entropy
of a
black hole could be computed  by counting the states of free strings. At the
time the focus was on Schwarzschild black holes. In order to make a
correspondence between the string and black hole states it was necessary to
postulate a large mass--renormalization of the string spectrum when the
coupling is turned on. While this mass shift is intuitively expected, it is
quantitatively difficult to compute. However, it was soon realized by Sen
[\sen] that the same logic could be applied to BPS black holes for which no
mass renormalization can occur. Since then the program of counting the states
of weakly coupled string theory and relating the degeneracy to BPS black hole
entropy has succeeded brilliantly [\stvaf] [\candj][\HorSt][\HorMldStr]. Here
we would like to
return
to the Schwarzschild case and describe a quantitative method for relating
strings and black holes.

Consider the degeneracy of a free (neutral) string at mass level $M^2=
8N_L=8N_R$
(where $\alpha^{\prime}=1/2$).
Standard methods give a degeneracy for large $N_{L,R}$,

$$d(N) = \exp{2\pi \left[\sqrt {N_L c_L \over 6}+\sqrt {N_R c_R \over
6}\right]}
\eqn\degen$$
where $c_{L,R}$ are constants equal to (24,24) for bosonic strings, (12,12) for
type II strings
and (12,24) for heterotic strings.  The entropy then satisfies

$$S_{string}= \log d(N) = 2 \pi \left[\sqrt {N_L c_L \over 6} + \sqrt{
 N_R c_R
\over 6}\right]
\eqn\stringentropy$$

On the other hand the Bekenstein-Hawking entropy of a Schwarzschild black
hole is given by

$$S_{BH} = 4 \pi M^2 G_N
\eqn\bhentropy$$

Obviously for large $M$ the quantum states of a \Sc \ black hole are much
denser than those of a free string at the same mass. In order to understand how
a correspondence can exist let us consider what happens to the free string when
the coupling constant $g$ is turned on. Obviously the mass of the
state begins to vary due to interactions. In particular the long range
gravitational interaction will begin to decrease the mass as the negative
potential energy increases. In fact no matter how small $g$ is, sufficiently
massive strings will undergo large gravitational corrections. For example a
string with level number satisfying $\sqrt N > g^{-4}$ will have a size smaller
than its \Sc \ radius and will certainly be subject to large corrections. Let
us then consider the evolution of the mass of the string state as $g$ is turned
up from zero to its final value. On very general grounds the mass levels will
be analytic functions of the parameter $g$. In general they will become
slightly complex since \bh s are unstable but the width of a typical \Sc \ \bh
\ is small, of order its inverse mass. In any case there should be no obstacle
to following the real part of the mass of a given state that begins at string
level $N$.
 It is obvious that the negative gravitational energy will
cause the levels to become more dense. If the levels become dense enough then
they can reproduce the level density implied by eq. \bhentropy. For example a
formula like

$$M^2 = {4N \over {l_s^2(1+\sqrt N g^2/(64\sqrt 2\pi))}}
\eqn\evolution$$
would turn the string degeneracy at $g=0$ into the black hole degeneracy
when
$\sqrt N g^2 >>1$ (see figure). However there does not seem to be much hope of
following
the
masses into the highly nonperturbative region of \bh s.

\let\picnaturalsize=N
\def\picsize{3.5in}
\def\picfilename{fig17n.ps}
\ifx\nopictures Y\else{\ifx\epsfloaded Y\else\input epsf \fi
\global\let\epsfloaded=Y
\centerline{\ifx\picnaturalsize N\epsfxsize \picsize\fi
\epsfbox{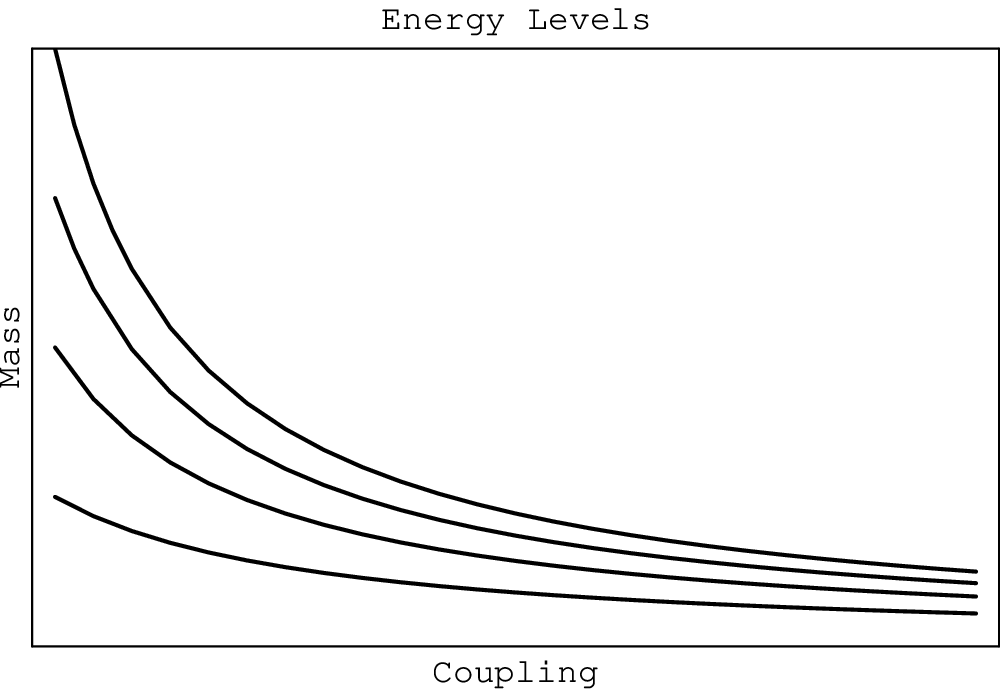}}}\fi
\centerline{ Evolution of the energy levels of a fundamental string as a
function the coupling constant  g.}

It is clear that almost all string states
lie within their \Sc \ radius and must evolve into \bh s as $g$ is turned on.
There is also a less well known argument that almost all uncharged black holes
evolved from
states of a single free string[\spec]. Consider what happens to a typical
uncharged black hole
in a large box  as $g$ is slowly turned off. It must evolve into some state
of
free string
theory  in the box, although not necessarily a single string (Other objects
such as D-branes and solitons become infinitely massive as $g \to 0$). However,
it has been known since the earliest days of string theory, that of the
states of
free string theory with a given mass the overwhelmingly most numerous are
single
strings.  Thus we expect that almost all black holes evolve back to single
string states. The purpose of this paper is to describe a strategy for
confirming
this hypothesis and that the Bekenstein--Hawking entropy just reflects the
level
density of the original strings.  Our
results are independent of the details of the
compactified six dimensional space. Furthermore they apply in all dimensions
less than or equal to 10.

%
%
\chapter{Strategy}

The strategy we will employ is the same one that has proved successful in
studying  D-brane systems in type II string theory just above extremality
[\candj] [\DMM] [\DMW]. In these references the properties of Hawking radiation
and absorption at  low energies were studied by using perturbation theory
around the weakly coupled limit in which black holes become perturbative
D-brane
systems. Although it is   not entirely clear why perturbation theory should
work, it seems that for wave lengths much longer than the Schwarzschild radius,
black holes behave like weakly coupled systems. In particular, the quantities
that have been successfully computed have the following features in common.
 First, they refer to very long wavelength. Second, when
expressed in terms of the entropy and discrete quantum numbers of the black
hole, the semiclassical expressions for these quantities are simple
positive powers of the coupling constant. An example of such a quantity is
the absorption cross section for scalar particles in the limit of
vanishing
frequency.

We begin with neutral systems. Let us assume that
 almost all \bh s
originate from single string states. In that case a given \bh \ can be labeled
by level numbers $N_{L,R}$ and its entropy
will be
$$S=2 \pi \sqrt {c \over 6} (\sqrt N_L + \sqrt N_R) \eqn\ent$$
We will consider
the independent quantity that specifies the \bh \ to be the entropy $S$. Now
suppose we are interested in the quantity $Q$ which can be computed in
semiclassical
\bh \ theory. In general, when expressed in terms of $S$ and $g$, $Q(S , g)$
will not be a power series but as we shall see certain quantities are. In this
case we can also hope to compute the same quantities in string perturbation
theory as a function of $\sqrt N$ and $g$. If the correspondence between black
holes and strings is correct the expressions should agree. Notice that this
strategy circumvents the need to calculate the mass shift.  As we
mentioned earlier, this strategy has only been tested for very low energy
quantities.

The quantity that we shall concentrate on is the area of the black hole.
Consider the semiclassical Bekenstein--Hawking relation

$$S={A \over 4G_N} \eqn\Bekhawk$$
Let us rewrite it as a formula for the area $A$.
$$
A=4G_N S =4\left({\kappa^2 \over 8\pi}\right) 2 \pi \sqrt {c \over 6} (\sqrt
N_L + \sqrt N_R)
\eqn\area
$$
 Evidently the area of the black hole is a perturbative quantity of order $g
^2$
and should be computable in string perturbation theory! This is a point that
has been emphasized by Maldacena[\Mld] in the context of BPS \bh s.

Now area is not one of the quantities that one normally thinks of computing
in
string perturbation theory. String theory is set up for the computation of
scattering amplitudes and decay rates. Therefore if we want to proceed we must
find an  expression for the area in terms of  on-shell matrix elements.
A number of possibilities come to mind. For example, the low energy limit of
 the
absorption cross
section for a massless particle to excite a black hole is known to be
proportional to the horizon area [\Unruh][\Gib]. In fact the  cross section
for a scalar particle at vanishing incident energy is exactly equal to the
horizon area[\GibDM].

A completely equivalent definition can be given in terms of the low energy
power spectrum  $P(\omega)$ of the Hawking radiation emitted by the black hole.
Unruh has calculated the power spectrum for massless minimally coupled scalars
and found for $\omega<<1/R_{BH}$

$$
P(\omega)={\omega^2 T \over 2 \pi^2} A_H
\eqn\power
$$
where $\omega $ is the frequency of the emitted quanta, $T$ is the temperature
of the emitter and $A_H$ is the horizon area. For our purposes we regard
\power\ as a definition of the area.
$$
A=
\lim_{{\omega}\rightarrow 0}{{2 \pi^2 P(\omega)}\over \omega^2 T}
\eqn\areadef
$$
Defined in this way $A$ is identical to the low energy limit of the absorption
cross section.

Our strategy will be to compute the temperature and the power spectrum of a
very highly excited string in powers of $g^2$.  From \area \ it follows that
if
we form the combination in \areadef  \ the higher orders beyond order $g^2$
should  vanish in the limit of large mass and the order $g^2$ term should
satisfy
\Bekhawk .  In this way we would derive the area of a black hole of entropy
$S$
from  the counting of levels of a quantum system. Exactly this type of
calculation has been successfully done in the D-brane theory of near extreme
black holes. [\DM][\DMM][\DMW].

%
%
\chapter{\Sc  \ Black Holes and Strings}

The temperature of a highly excited weakly coupled neutral string is easy to
compute to
leading order in perturbation theory. The entropy of the free string is
proportional to its mass $M$. Using eq. (2.1) the first law  gives

$$
\beta={1 \over T}={dS \over dM}={\pi \over 2} \left[\sqrt {c_L \over 3} +\sqrt
{c_R \over 3} \right]
\eqn\hag
$$
This is just the Hagedorn temperature at which a very weakly coupled string
will radiate.
Although we have not calculated the perturbative corrections to the temperature
there is no reason for them to be absent. Thus the temperature of a string at
large level number should have a perturbation expansion of the form

$$
T = T_{Hagedorn} -{g^2 }F(N) +...
\eqn\texpans
$$

The luminosity $P(\omega)$  is more complicated and will be calculated in terms
of decay
rates. Obviously the decay rates and therefore $P(\omega)$  vanish for $g=
0$.
The
leading term is order $g^2$. Therefore when calculating the area to order $g
^2$
we only need the temperature to leading order.

A classical black hole solution represents a statistical ensemble of states.
The initial state of  a free string that  we wish to consider  should also
 be a
statistical  ensemble  defined by introducing a thermal density matrix which
 is
peaked at states with the desired mass. Recalling the first quantized
expression for the mass.

$$
M^2 =8 N_L = 8 N_R = 4 N
\eqn\masssquare
$$
we are led to a density matrix of the form
$$
\rho = Z^{-1} exp(-\beta_L^* N_L-\beta_R^* N_R)
\eqn\density
$$
where $Z$ is defined so that $Tr \rho = 1$. It should be noted that $\beta^
*$
is not the inverse of the real temperature of the system. It is a
dimensionless parameter, which
is chosen to
fix the average value of $M^2$. One  finds that
$$
\beta_{L,R}^* ={dS \over dN_{L,R}}= \sqrt {\pi^2 c_{L,R} \over 3N_{L,R}}
\eqn\bet
$$

Now consider the emission of a scalar particle by a typical member of
the ensemble. Let us choose the particular scalar that corresponds to the
component
$g_{56}$ of the graviton. The vertex operator for this scalar is (using the
conventions of [\DM])

$$
V(k) = \int {4 \sqrt {2} \kappa \over \pi}[\pa_+ X^5 \pa_- X^6  + \pa_- X^5
\pa_+ X^6  + fermion \quad
terms]e^{ikX}d^2 \sigma
\eqn\vertex
$$
where the derivatives refer to world sheet light cone coordinates and the
momentum $k$ is a null vector in the four dimensional  uncompactified Minkowski
space. If the momentum $k$ is much smaller than $l_s^{-1}$ in the rest
frame of the
decaying string then  the fermionic term  and  the factor $e ^{ikx}$   in the
vertex operator
can be ignored except for the center
of mass contribution  which when integrated out provides a momentum conserving
delta function.

The usual oscillator representation for the $X's$ leads to the
expressions

$$
\pa_+ X^{\mu} = \sum \alpha^{\mu}_n e^{-2in \sigma_+}
$$
$$
\pa_- X^{\mu} = \sum {\tilde\alpha}^{\mu}_n e^{-2in \sigma_-}
\eqn\prtx
$$

The matrix element for the decay of a state $|i \rangle$ which we take to be
at rest,  to a state $|f \rangle$  by colliding a right moving $X^5$ with a
left moving $X^6$ and emitting a scalar $g_{56}$ has the form

$$
{\cal M}={4 \sqrt 2 \kappa \over \pi}
\langle i|\int \sum_{n,m} \alpha^5_n \tilde \alpha^6_m
e^{-2i(n-m)\sigma}d\sigma
|f \rangle \delta^4(p_i+k-p_f)
\eqn\amp
$$
where the vertex function no longer contains
the factor $e^{ikX} $. The delta function constrains the on-shell momenta of
the initial and final string. An analogous expression comes from the second
term in \vertex. In practice, if the mass of the initial and final
strings are much larger than the energy $\omega$ carried by the scalar then
the
only effect of the delta function is to constrain the masses according to

$$
M_i = M_f + \omega
\eqn\energy
$$

Let us assume that the initial and final strings are at levels $N$ and
$N-\delta N$. Then using $M^2=4N$ we find

$$
\omega=\delta M={2n \over M}
\eqn\delt
$$
where $n=\delta N$. The process of decay is now seen to have a simple intuitive
structure. The
vertex operator may be averaged over the world sheet coordinate $\sigma$ and
becomes

$$
V ={4 \sqrt 2 \kappa} \sum (\alpha^5 _n \tilde \alpha^6_n + \tilde \alpha^5 _n
\alpha^6_n)
\eqn\vet
$$
It describes the annihilation of two oppositely moving quanta on the string
with
mode number $n$. The energy is carried off by the scalar whose energy is
constrained to satisfy eq. \delt.

To obtain the decay rate per $d\omega$ we square the amplitude,
average over the initial thermal distribution and
multiply by the density of resonances $M/4$ \delt

$$Tr \rho \sum_f |{\cal M}|^2=32 \kappa^2 n^2 {1 \over
(e^{\beta_L^*n}-1)(e^{\beta_R^*n}-1)} 2 ({1\over 2M})^2 {M \over 4}
\eqn\temp$$
where the factor $2$ comes from the two terms in \vet,
 and $(1/2M)^2$ from the relativistic normalization of the initial and final
states.
The luminosity $P(\omega)$, in the low frequency limit, is given by Fermi's
golden rule to be
$$P(\omega)={8 \kappa^2 \over M} {\omega^2 \over {2 \pi \beta^*_L \beta^*_R}}
\eqno
(3.13)$$
where we have used eq. (3.12) and the fact that $\beta_{L,R}^*n<<1$.

This has to be compared with the classical result for luminosity which we
use as a definition of the area of the black hole horizon
$$P(\omega)= {\omega^2 \over {2 \pi^2 \beta}} A_H \eqno (3.15)$$
from which we find
$$ A_H= {64\pi^2\beta G_N \over M\beta^*_L \beta^*_R}$$
where we have used  $\kappa^2=8 \pi G_N$.

The value of $\beta$ to be used in (3.15) is the lowest order expression given
by eq. (3.5).
 Expressing $\sqrt N$ in terms of the
entropy,
from eqs. (3.14) and (3.15) we get
$$4G_N S{\sqrt{36 \over {c_Lc_R}}}=A_H \eqno (3.16)$$

Two things are apparent from eq (3.16). The first is  that the entropy and area
are indeed  proportional.  Moreover,  as we will see in the next section, this
proportionality extends to black holes in arbitrary dimension as well as
arbitrary charges and angular momentum. The  numerical proportionality factor
is always the same.

The second point is that the proportionality factor is not the correct
Bekenstein Hawking factor unless $c_L c_R=36$.  Unfortunately this value does
not correspond to to any fundamental string theory. The meaning of this result
is very unclear but it suggests that there may be some kind of nonperturbative
"renormalization" of $c$ in the environment of a horizon .

\chapter{Charged and Rotating Black Holes }

We can extend the above calculations to charged and rotating black holes.  The
classical
absorption cross section for charged black holes in arbitrary dimensions was
calculated in
[\GibDM] and was shown to be equal to the area. The absorption cross section
for a four
dimensional Kerr black hole was calculated in [\Ford] and was also shown to be
equal to the area.
We will assume that this result generalizes to arbitrary charged rotating black
holes and so
the classical relation (3.15) continues to hold.

In general we can write
$$S=S_L+S_R\eqno (4.1)$$
where $S_L^2$ is linear in $N_L$ and $S_R^2$ is linear in $N_R$. For example
for charged strings we have
$$S_L^2=\left( 2\pi\sqrt {c_L \over 6}\right)^2N_L\eqno (4.2)$$ and for
rotating bosonic
strings with $J\sim N_L$
$$S_L^2=\left( 2\pi\sqrt {c_L \over 6}\right)^2(N_L-J)\eqno (4.3)$$
with similar  equations for $S_R$.
Also  $M^2=8N_L+Q_L^2=8N_R+Q_R^2$ .

Now we can repeat the calculation of the previous section. The only changes
are in the $\beta^*_{L,R}$ and the expressions for $M$ and $S$.
In particular, we find
$$ A_H=({64\pi^2G_NS })({\beta \over M\beta^*_L\beta^*_RS})\eqno (4.4)$$

We will evaluate the expression in the second parenthesis and show that it is
independent of the charges and angular momentum if and only if $c_L=c_R$. The
relation between area and
entropy will therefore be identical to the one in the previous section.
We have
$$\beta^*_{L,R}={dS_{L,R} \over dN_{L,R}}={\left( 2\pi\sqrt {c_{L,R} \over
6}\right)^2  \over 2 S_{L,R}}\eqno (4.5)$$
$$\beta={dS \over dM}={dS_L \over dM}+{dS_R \over dM}\eqno (4.6) $$
Now
$${dS_L \over dM}=({dS_L \over dN_L})({dN_L \over dM})={\left( 2\pi\sqrt
{c_{L,R} \over 6}\right)^2 M \over 8S_L}\eqno (4.7)
$$
and so
$$\beta=M({\left( 2\pi\sqrt {c_L \over 6}\right)^2 \over 8S_L}+{\left(
2\pi\sqrt {c_R \over 6}\right)^2 \over 8S_R})\eqno (4.8)
$$
If we now evaluate the expression in parenthesis in eqn(4.4) we find that it is
independent
of all charges and angular momenta only if $c_L=c_R=c$. For this case we find
$${\beta \over M\beta^*_L\beta^*_RS}= {1 \over 2\left( 2\pi\sqrt {c \over
6}\right)^2}\eqno (4.9)
$$
Thus, as we anticipated,  the relation between the entropy and area is
unaffected
by the presence of charges and angular momentum for $c_L=c_R=c$. We therefore
reproduce the
Bekenstein-Hawking entropy for $c=6$ as in the previous section.

The generalization to arbitrary dimensions is a consequence of the independence
of
the string scattering amplitude on dimension and the equality of the phase
space factors
for the string and the classical calculation. So the correspondence of area and
entropy
should hold in any dimension.

\chapter{Conclusions}

We have seen that the low energy absorption cross section $\sigma$ for scalars
from
fundamental  strings satisfies
$$\sigma_{abs}= 8\pi G_N \sqrt {cN \over 6} $$
Combining this with a fact and an assumption leads to  a proportionality
between entropy and area. The fact is that the absorption cross
section at
$\omega \to 0$ for any  black hole is the area of the
classical
horizon. The assumption is that the levels of a free  string are
in one to
one correspondence
with the levels of  black holes that evolve from the strings as $g$ is
increased.
If this is true, we may replace $\sqrt{cN \over 6}$ for the string with the
entropy of the
black hole ensemble.
Thus, the assumption leads to a relation between entropy and area.
However, the precise  perturbative  calculation  of the numerical
proportionality factor does not agree with the Bekenstein-Hawking  value
unless $c=6$. We do not understand the meaning of this but it suggests that
there may be some kind of ``renormalization'' of $c$  in the environment of a
black hole.

\chapter{Acknowledgements}
We would like to thank  G.W. Gibbons, S.R. Das and
G.Mandal for  discussions. We would also like to thank J. Maldacena, S. D.
Mathur and
S. de Alwis for discussions and especially for pointing out that  agreement
with the Hawking Bekenstein formula requires $c=6$.
The work of B.K and L.S. is supported by NSF grant PHY-9219345. The work of
A.R is supported in part by the Department of Energy under contract no.
DE-AC03-76SF00515.

\refout
\end